\documentclass[aps,prb,reprint,twocolumn,superscriptaddress,showpacs,floatfix]{revtex4-1}

\usepackage{graphicx}
\usepackage{amsmath}
\usepackage{amssymb}
\usepackage{bbm}
\usepackage{color}
\usepackage{enumitem}
\usepackage{nicefrac}
\usepackage{multirow}
\usepackage{braket}

\usepackage[colorlinks,citecolor=red,urlcolor=blue,bookmarks=false,hypertexnames=true]{hyperref} 

\begin{document}

\title{Improving excited state potential energy surfaces via optimal orbital shapes}
\date{\today}
\author{Lan Nguyen Tran}
\email{lantrann@berkeley.edu}
\affiliation{Department of Chemistry, University of California, Berkeley, California, 94720, USA}
\affiliation{Ho Chi Minh City Institute of Physics, VAST, Ho Chi Minh City 700000, Vietnam}
\author{Eric Neuscamman}
\email{eneuscamman@berkeley.edu}
\affiliation{Department of Chemistry, University of California, Berkeley, California, 94720, USA}
\affiliation{Chemical Sciences Division, Lawrence Berkeley National Laboratory, Berkeley, CA, 94720, USA}

\begin{abstract}
We demonstrate that, rather than resorting to high-cost dynamic correlation methods,
qualitative failures in excited state potential energy surface predictions
can often be remedied at no additional cost by ensuring that optimal molecular
orbitals are used for each individual excited state.
This approach also avoids the weighting choices required by
state-averaging and dynamic weighting and obviates their need for
expensive wave function response calculations when
relaxing excited state geometries.
Although multi-state approaches are of course preferred near conical
intersections, other features of excited state potential energy surfaces
can benefit significantly from our single-state approach.
In three different systems, including a double bond dissociation,
a biologically relevant amino hydrogen dissociation, and an
amino-to-ring intramolecular charge transfer, we show that state-specific
orbitals offer qualitative improvements over the state-averaged status quo.
\end{abstract}

\maketitle

{\it Introduction.}
Exited-state geometry relaxations are an essential
phenomenon in molecular photochemistry.
Alongside inter-state properties like non-adiabatic couplings and
transition dipoles, single-state potential energy surface (PES)
features like the depths and locations of excited-state minima and
the heights of barriers between minima help determine how a molecule
will respond and transform under exposure to light.
For example, selecting ligands in order to extend the lifetimes of
Fe-based metal-to-ligand charge-transfer (MLCT) states depends
centrally on the relative energies and positions of the MLCT
and other excited state minima.
\cite{deGraaf2010,deGraaf2011,zhang2017manipulating,wenger2019iron,wu2020controlling}
While density functional theory (DFT), with its low cost and often excellent
ground state energetics, is widely used in studying excited states,
its well-known difficulties with charge-transfer states and in systems
exhibiting strong electron correlations make reliable, low-cost alternatives
a high priority in the study of photochemistry.
The complete active space self-consistent field (CASSCF) approach 
is often a powerful alternative,
\cite{gonzalez2012,MOLCAS2013,serrano2005}
but it is significantly more difficult to use than DFT, and, due to its own
limitations, can still make qualitative errors in PESs without the aid
of expensive post-CASSCF correlation corrections.
A good example of this frustrating reality occurs in the charge transfer
state of 4-aminobenzonitrile (ABN), where state-averaging compromises
prevent the wave function from undergoing proper orbital relaxation and
lead CASSCF to predict a qualitatively incorrect excited state geometry.
\cite{segado2016intramolecular} 
Although applying corrections with
complete active space perturbation theory (CASPT2) brings
predictions in line with experiment, \cite{segado2016intramolecular} 
such post-CASSCF correlation methods greatly increase computational cost.
While such methods' incorporation of the finer details of
electron correlation is essential for high-precision energetics,
we show here that in this case and others, qualitatively correct predictions
can be restored simply by making the molecular orbital shapes optimal for
individual excited states.

%Further, we will show that in this case, as well as in other cases where
%CASSCF suffers qualitative failures for excited state PESs,
%improving the quality of excited state molecular orbitals, and in particular
%making them optimal for the excited state in question,
%both reduces the complexity of CASSCF geometry relaxations and restores
%qualitatively correct predictions without having to resort to
%expensive post-CASSCF methods.
%Thus, although post-CASSCF corrections are certainly valuable,
%it appears that  using optimal orbital shapes for excited states,
%a luxury long enjoyed by ground states, can greatly
%simplify and accelerate the study of many aspects of excited state PESs.

While the widely used practice of state averaging (SA) has seen many
successes in excited state investigations, it faces
a number of important challenges.
In the SA approach, the molecular orbital shapes are chosen by minimizing
the weighted average of multiple states' energies.
This appears at first glace to be balanced, and in many cases is, but
can also entrench imbalances if the needs of one state (e.g.\ strong
orbital relaxations following charge transfer) are denied in
favor of the needs of others (e.g.\ multiple local excitations that
should not involve strong orbital relaxations).
As we discuss below, this problem appears to be responsible for
SA-CASSCF's failure to predict an excitation-induced twist in ABN.
SA-CASSCF is also a popular way to address the challenge of root flipping,
where an optimization fails to converge due to two states exchanging
back and forth in the energy ordering.
However, SA is not a panacea here, as there is always the risk of the
highest-energy state in the average flipping with the next state not
included in the average.
Glover, for example, has recently shown explicitly that SA-CASSCF
still suffers from root flipping. \cite{Glover2014}
%While single-reference correlated approaches like density functional theory (DFT), perturbation theories, and coupled-cluster methods are affordable and accurate for weakly correlated systems, they usually fail for cases involving strong correlations that can be essential for correctly describing many real molecules like MLCT complexes. 
%Complete active space self-consistent field (CASSCF) theory is powerful for such problems. State-averaging CASSCF (SA-CASSCF) is usually used to treat excited states to avoid the root-flipping problem present in state-specific CASSCF (SS-CASSCF). Another important feature of SA-CASSCF is that it can facilitate multi-state calculations, such as, state interactions and non-adiabatic couplings needed for conical-intersection optimization and non-adiabatic dynamics simulations. However, as recently shown by Glover \cite{Glover2014}, SA-CASSCF still encounters the root-flipping problem. Therefore, as will be shown later in this paper, expected states may not be seen within a given number of states in averaging, leading to an incompleteness of PESs.
Another challenge is that SA-CASSCF often introduces discontinuities
in potential energy surfaces (PESs),
\cite{deskevich2004,Glover2014,Glover2019,tran2019tracking}
which can be a particular problem for dynamics simulations.
\cite{Glover2014,hollas2018}
Whether these discontinuities arise from the highest state in the
average crossing the lowest state not in the average as the atoms move
around, or from a cusp in a single state being converted to a discontinuity
in all states via the SA link,
this high-priority problem has attracted a good deal of attention.
%Numerous efforts have been devoted to bypass issues present in SA-CASSCF.
One approach to address the issue is to prepare good orbital sets for
complete active space configuration interaction (CASCI), such as 
floating occupation molecular orbitals (FOMO), \cite{slavivcek2010ab}
and then forgoing the SA-CASSCF orbital optimization entirely.
While less prone to PES discontinuities, this approach unfortunately can
still produce them for high-lying excited states. \cite{yang2018imaging}
An alternative approach, and one that is particularly relevant near conical
intersections where retaining at least a two-state treatment is
advantageous, is to weight the SA dynamically so as to favor the needs
of the important states while retaining the stability offered by SA.
\cite{deskevich2004,Glover2014,Glover2019}
The drawback is that the now energy-dependent weights complicate the
evaluation of analytic gradients even more than SA does already.
\cite{Glover2014,Glover2019}
Ideally, discontinuities and root flipping would be avoided
while retaining ground-state CASSCF's simple analytic gradients,
each state would have the orbital relaxations it needs,
and the user would not need to make any decisions about weightings
that affect the final predictions.

Towards these ends, we present in this paper a study of how
fully excited-state-specific CASSCF, in which the molecular orbitals
are optimized solely for the benefit of the state in question and
collapse to lower states is avoided through careful state-tracking
methods, \cite{tran2019tracking} can achieve these goals and succeed
in cases where SA-CASSCF suffers qualitative failures.
In addition to the case of ABN, where experiment and expensive
post-CASSCF methods predict a twisted charge transfer geometry while
SA-CASSCF does not, \cite{grabowski2003,misra2018intramolecular}
we investigate the crossing of analine's first $^1\pi\pi^*$ and
charge transfer states (which SA-CASSCF fails to predict
\cite{sala2014new,ray2018conical})
and PES disappearance and discontinuity in thioacetone.
In addition to qualitative improvements in accuracy, it is important
to emphasize that the excited-state-specific approach, by making
the state's energy stationary with respect to the orbital shapes,
avoids the response calculations that SA-CASSCF and dynamic-weighting
need to perform \cite{yamaguchi1994} when evaluating analytic gradients.
Avoiding the response evaluations matters, as they are
more expensive \cite{Granovsky2015} and can be more prone to
convergence issues \cite{dudley2006parallel} than CASSCF itself.
In our approach, analytic gradients are no more difficult
than in the ground state case, and indeed can use the exact same
gradient code.
While these advantages in both accuracy and numerical simplicity are
exciting, we should keep in mind that SA-CASSCF is popular
precisely because it is often successful, and so it is important to
emphasize the particular situations in which it is most in need of
assistance.
In this regard, we will point to charge transfer, which is a
technologically important case where SA is at particular risk of
entrenching biases between states rather than creating balance.
\cite{domingo2012metal,pineda2018excited,tran2019tracking}

{\it Theory.}
Our state-specific approach (SS-CASSCF) rests on the general property that
exact Hamiltonian eigenstates are energy stationary points.
In the context of an approximate ansatz like CASSCF, the idea is to find
the ansatz's stationary point that corresponds to the
excited state under study.
For simplicity, we retain the two-step approach common to many CASSCF
implementations in which the orbitals and CI variables are optimized
separately.
In particular, the orbitals are optimized so as to minimize the energy
gradient norm \cite{tran2019tracking}
(rather than the energy, which would encourage collapse
towards the ground state), while the CI variables are chosen at each
stage as the CASCI root most similar to the desired state.
While there are of course many ways to make a precise definition for
what one means by similar, we have found \cite{tran2019tracking}
that the measure
\begin{align}
\label{eqn:qwg}
Q_{W\Gamma}
%(\vec{c}\hspace{0.5mm})
&=
%   W(\vec{c}\hspace{0.5mm})
% + D(\vec{c}\hspace{0.5mm}),
%W(\vec{c}\hspace{0.5mm}) =
\frac{\braket{\Psi|(\omega-\hat{H})^2|\Psi}}{\braket{\Psi|\Psi}}
+
%\label{eqn:ddm}
%D(\vec{c}\hspace{0.5mm}) &=
 \frac{\hspace{1mm} || \hspace{.5mm} 
       \Gamma_t - \Gamma%(\vec{c}\hspace{0.5mm})
       \hspace{.5mm} || \hspace{1mm}}
       {n_{{}_{\mathrm{CAS}}}}
\end{align}
is particularly effective.
Here $\omega$ is a target energy that we think is close to the
energy of the state we are after, $\Gamma$ is our state's one-body
density matrix, $\Gamma_t$ is a target density matrix (taken either
from an initial CASCI or from the previous iteration), and
$n_{CAS}$ is the number of active orbitals.
The idea is that, after finding the low-energy roots of the CASCI
problem, we select the root with the lowest value for $Q_{W\Gamma}$
and then perform an orbital optimization step that seeks to make
that root's energy stationary with respect to orbital changes.
Crucially, the selection of the root based on its energy and
density matrix makes the approach insensitive to root flipping,
so even in cases where the order of the states in the CASCI changes
as the orbitals are optimized, the approach converges
to the energy stationary point corresponding to the desired
excited state. \cite{tran2019tracking}
Note especially that the precise choices for $\omega$ and $\Gamma_t$
do not affect the final outcome, so long as they lead the optimization
to converge to the correct CASSCF stationary point.
In other words, the final energy depends only on the wave function at
the energy stationary point, which is independent of
$\omega$ and $\Gamma_t$, but made easier to find and converge to via
intelligent choices for these parameters.

Once the energy is stationary, the Hellman-Feynman theorem guarantees
that analytic gradients with respect to nuclear coordinates $R$
simplify to
\begin{align}
    \frac{dE}{dR} = \left<\Psi\right|\frac{\partial \hat{H}}{\partial R}\left|\Psi\right>
\end{align}
in a direct parallel to the situation for ground states.
There is no need for response calculations, as full energy
stationarity eliminates the wave function response terms from the gradient
expression.
In SA-CASSCF, in contrast, the energies of individual states that control
the PES are not stationary (it is only their weighted average that is)
and so additional terms involving the wave function's response to the
geometry distortion appear in the gradient equation and must be evaluated.
%The wavefunction $\left|\Psi\right>$ is optimized specifically using our recently-introduced $W\Gamma$ root tracker \cite{tran2019tracking} that is a composite of excited-state variational principle \cite{messmer1969variational,messmer1970variational} and density matrices
%\begin{align}
%\label{eqn:qwg}
%Q_{W\Gamma}(\vec{c}\hspace{0.5mm}) &=
%   W(\vec{c}\hspace{0.5mm})
% + D(\vec{c}\hspace{0.5mm}),
%\end{align}
%with
%\begin{align}
%W(\vec{c}\hspace{0.5mm}) = \frac{\braket{\Psi|(\omega-\hat{H})^2|\Psi}}{\braket{\Psi|\Psi}},
%\end{align}
%and
%\begin{align}
%\label{eqn:ddm}
%D(\vec{c}\hspace{0.5mm}) &=
% \frac{\hspace{1mm} || \hspace{.5mm} 
%       \Gamma_t - \Gamma(\vec{c}\hspace{0.5mm})
%       \hspace{.5mm} || \hspace{1mm}}
%       {n_{{}_{\mathrm{CAS}}}}. 
%\end{align}
%
%In the first term, eigenstates are targeted by the parameter $\omega$ that is dynamically updated. To reduce computational costs, an approximation to the first term has been usually used. Details of derivation and implementation are given in our previous paper \cite{tran2019tracking}. In the second term, states' properties are targeted by matching one-particle density matrices. $\Gamma_t$ is the targeted density matrix and rotated to current orbitals at each macro iteration. $n_{CAS}$ is the number of active orbitals.
Our analytic gradients in hand, we perform our geometry relaxations
using the {\it geomopt} module within pySCF, \cite{pyscf2018}
which interfaces with geomeTRIC \cite{wang2016geometry} and
PyBerny \cite{pyberny} for constrained and unconstrained geometry
optimizations, respectively.
Comparison calculations with SA-CASSCF were performed using Molpro.
\cite{werner2012molpro}
%To track desired states between different geometry optimization loops, we use the density matching criterion (\ref{eqn:ddm}).
%by connecting our modified pySCF \cite{pyscf2018}.
%In the current implementation, we have employed the two-step approach of switching back and forth
%between Davidson CI diagonalizations and orbital optimization steps alternatively.
%The ground-state orbital optimization that minimizes energy can lead to variational collapse.
%Therefore, we optimize orbitals using a modified Newton-Rapshon method in which the generalized minimal residual (GMRES) method followed by a line search set to minimize $|\nabla E|^2$ rather than $E$ itself is used. 

\begin{figure}[t]
  \includegraphics[width=10cm,]{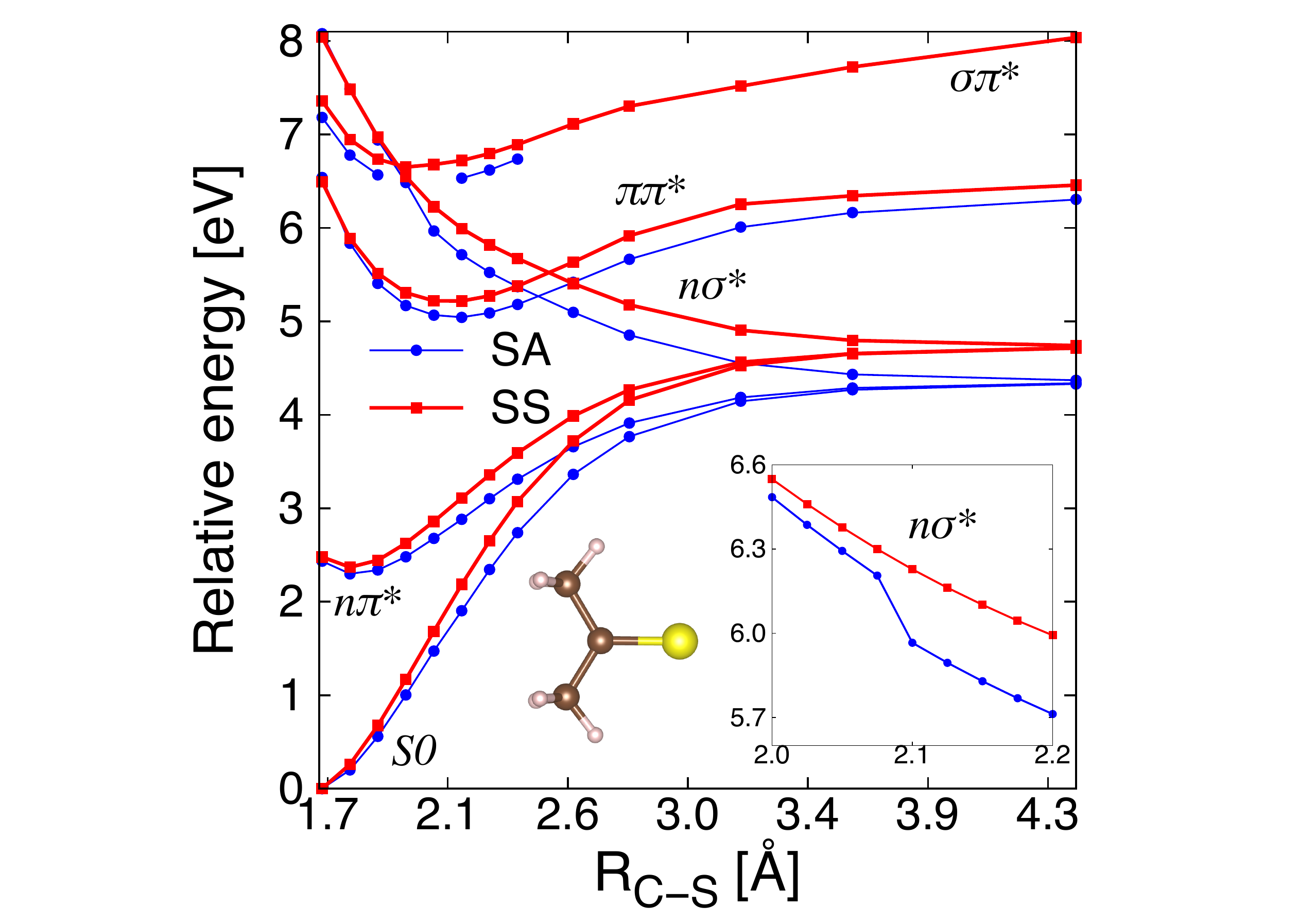}
  \caption{
  Potential energy curves showing, for each low-lying singlet state of
  thioacetone, the energy of the state after relaxing the geometry
  with the C--S bond distance $R_{C-S}$ held fixed.
  SA-CASSCF used a 5-state SA with equal weights, while both SA-CASSCF
  and SS-CASSCF employed the 6-31G(d) basis and a (6e,5o) active space
  containing the $\sigma$, $\sigma^*$, $\pi$, $\pi^*$, and lone pair $n$ orbitals.
  }
  \label{fig:c3h6s}
\end{figure}

{\it Results.}
Let us begin with the relatively simple example of C--S bond
photodissociation, which can be important in
astrochemistry, \cite{xu2019ab}
biomedicine, \cite{mao2015photoinduced}
and catalysis. \cite{ganguly2020comparative}
For C--S single bond dissociation, non-radiative internal conversion between
the first excited state $^1\pi\pi^*$ and the dissociative $^1\pi\sigma^*$
state is key. \cite{ashfold2017exploring}
Here we instead investigate the C--S double bond dissociation of thioacetone,
(CH$_3$)$_2$CS, using both SA-CASSCF and SS-CASSCF.
By performing geometry relaxations for each of the low-lying singlet states
at a series of fixed C--S bond distances, we can see
which states are expected to be key for the double bond dissociation.
We see in Figure \ref{fig:c3h6s} that, among the low-lying singlets,
it is the $^1n\sigma^{*}$ state that is dissociative in character,
while the $^1n\pi^*$, $^1\pi\pi^*$, and $^1\sigma\pi^*$ surfaces are all bound.
For the lowest four states, the SA and SS approaches are in qualitative
agreement, but SS-CASSCF displays a clear advantage for the $^1\sigma\pi^*$ state.
For this state, the SA-CASSCF geometry optimization was unable to converge at
most C--S bond distances, while SS-CASSCF optimizations converged at all distances.
While it is possible that extending the SA to include more than five states could
help here, this is highly undesirable, as including more states leads to
each state having even less say in how the orbitals should be shaped, and, worse,
increases the chances of finding discontinuities, as a discontinuity in any states'
surface gets spread to all states through their link in the SA energy.
With five states in the average, SA-CASSCF suffers one discontinuity already, which
the inset of Figure \ref{fig:c3h6s} shows near the $R_{C-S}=2.1$\r{A} geometry of
the $^1n\sigma^*$ state.
(Note that, although at that geometry all states show a discontinuity due to the
SA link, this is not seen in the figure, as the other states' energies are being
reported at their own relaxed geometries).
The discontinuity is relatively small, and does not alter the basic 
dissociative character of the state, but even small discontinuities can
strongly affect dynamics simulations. \cite{Glover2014,Glover2019}
In the SS approach, in contrast, the $^1\sigma\pi^*$ and $^1n\sigma^*$ states
are quite well behaved, and we need not agonize over how many states to include
in any average and how this may affect orbital quality.
With the molecular orbital shapes now optimal for each state individually,
these difficulties are avoided, the geometry optimizations all converge
successfully, and no discontinuities are encountered.

We now turn our attention to aniline, (C$_6$H$_5$NH$_2$), a common
basis unit for biomolecules whose N--H photodissociation is important
for medical applications, including UV radiation protection within sunscreen.
\cite{roberts2014role}
Aniline's excited-state dynamics have been extensively studied,
\cite{king2010dynamical, roberts2012unraveling, wang2013quantum,
      roberts2014role,  sala2014new, ray2018conical, jhang2019triplet}
and, as is common for heteroaromatic biomolecules, the $^1\pi\sigma^*$
state has been shown to facilitate the photodissociation process.
\cite{roberts2014role}
It is particularly noteworthy that this state is
quasi-bound in the region of the S0 minimum, with dissociative character
only appearing once the N--H bond has been stretched.
Equation of motion coupled cluster (EOM-CCSD) predicts the barrier to
leave the quasi-bound region and dissociate to be 0.5 eV, \cite{wang2013quantum}
%Therefore, the quasi-bound and dissociative components of the $^1\pi\sigma^*$ PES are separated by a barrier estimated at 0.5 eV using EOM-CCSD \cite{wang2013quantum}.
and experimental investigations suggest that the first $^1\pi\pi^*$ excited state
intersects the $^1\pi\sigma^*$ state near the latter's quasi-bound local minimum
geometry. \cite{king2010dynamical}
Previous theoretical investigation has failed to predict this crossing
using SA-CASSCF, \cite{sala2014new, ray2018conical}
%Regarding theoretical calculations, some previous authors could not see this crossing at the SA-CASSCF level \cite{sala2014new, ray2018conical}.
and it is only when expensive post-CASSCF methods like
extended multi-configuration quasi-degenerate second-order perturbation theory
\cite{sala2014new} (XMCQDPT2) and extended multi-state multi-reference
perturbation theory \cite{ray2018conical} (XMS-CASPT2) are employed that
a crossing is predicted.
With the $^1\pi\sigma^*$ state showing a substantial change in dipole moment
compared to the ground state and thus possessing at least some charge transfer
character, we wondered whether predicting the crossing really required such
expensive methodology or, instead, whether this was a case in which states with
substantially different needs in terms of molecular orbital shapes were being
ill-served by the compromises inherent to state averaging.

\begin{figure}[t]
  \centering
  \includegraphics[width=8cm,]{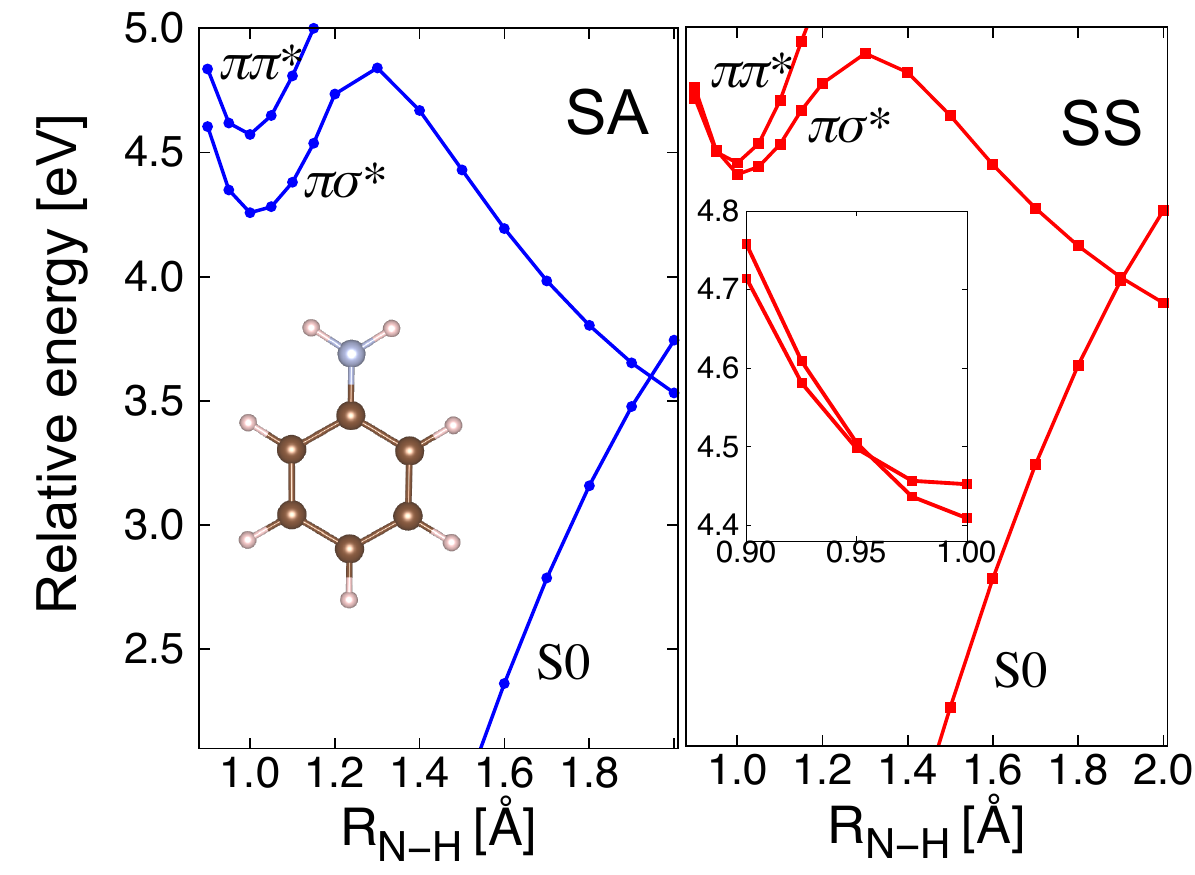}
  \caption{
  Potential energy curves showing, for three low-lying singlet states
  of aniline, the energy of the state after relaxing its geometry under
  $C_s$ symmetry with the N--H bond distance $R_{N-H}$ held fixed.
  %The right inset shows a zoom in the range $0.9-1.0\r{A}$
  %for the $^1\pi\pi^*$ and $^1\pi\sigma^*$ states.
  SA-CASSCF used a 3-state SA with equal weights, while both SA-CASSCF
  and SS-CASSCF employed the 6311++g(d,p) basis and a (10e,9o)
  active space containing seven $\pi$/$\pi^*$ orbitals,
  an N--H $\sigma$ orbital, and an N--H $\sigma^*$ orbital.
  }
  \label{fig:anl_pes}
\end{figure}

\begin{figure}[t]
  \includegraphics[width=8cm,]{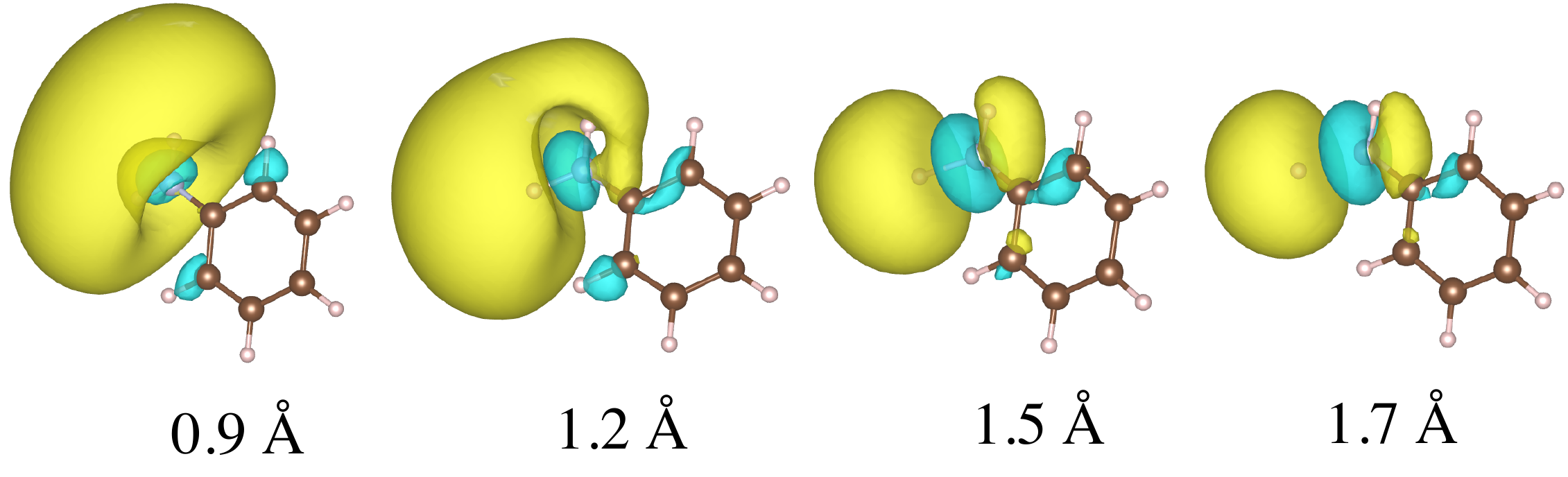}
  \caption{
  The $\sigma^*$ orbital of aniline in the SS-CASSCF $^1\pi\sigma^*$ state
  at various N--H bond distances.
  }
  \label{fig:anl_orbs}
\end{figure}

As seen in Figure \ref{fig:anl_pes}, our SA-CASSCF calculations also fail
to predict a crossing, whereas SS-CASSCF predicts a crossing near an N--H
bond distance of 0.95 \r{A}.
Although the crossing is predicted by post-CASSCF methods to occur
closer to a bond distance of 1.05 \r{A}, showing that
post-CASSCF correlation effects
do play a quantitative role here, the fact that SS-CASSCF predicts
a crossing at all is a qualitative improvement over the SA approach.
We also note that, although it is a smaller advantage,
the SS-CASSCF dissociation barrier of 0.49 eV is
closer to the 0.5 eV EOM-CCSD prediction \cite{wang2013quantum}
than is the 0.58 eV barrier of SA-CASSCF.
To understand why SA-CASSCF has difficulty here, we have plotted
the $\sigma^*$ orbital from our SS-CASSCF $^1\pi\sigma^*$ state in
Figure \ref{fig:anl_orbs}, where we see that the charge transfer
character suggested by the large dipole change is really a Rydberg-like
extension of the $\sigma^*$ orbital off one side of the molecule.
The large change in dipole is nonetheless present, which creates a
state-specific need for the non-active orbitals to respond and
re-polarize their electron distributions.
This effect is difficult to achieve in SA-CASSCF, as these re-polarizations
are inappropriate for the other two states.
We find that, even if we us a biased S0/$\pi\sigma^*$/$\pi\pi^*$ weighting
of 20/40/40 and re-optimize the two excited state geometries with
$R_{N-H}$ fixed at 0.95 \r{A}, the excited state energies do not get
closer together.
This finding strongly suggests that dynamic weighting would not help here,
which is not surprising as, again, the issue is that the two states have
significantly different dipoles and so require different post-excitation
orbital relaxations.
By instead going in for fully state-specific orbitals, which can be seen as the
logical endpoint of biased weighting, the different states get the
orbital relaxations that are appropriate to them, and a crossing
is successfully predicted.

%A 20/40/40 biased weighting in SA did not move the energies
%closer together at the 0.95 geometry \AA, nor when we
%re-optimized the geometry at this distance with the biased
%weighting, which suggests that it may be challenging for
%dynamical weighting schemes to observe this crossing.

%Analyzing dipole moments, we see that the state $^1\pi\sigma^*$ has a strong charge-transfer character.
%Therefore, in this work, we have attempted to see the $^1\pi\pi^*/\pi\sigma^*$ crossing using our SS-CASSCF implementated in pySCF.
%We have employed the Molpro package to perform SA(3)-CASSCF involving the S0, first $^1\pi\pi^*$, and $^1\pi\sigma^*$ states with equal weights.

%We have used the 6311++g(d,p) basis set. An active space of 10 electrons in 9 orbitals including seven $\pi$, $\pi^*$ orbitals, and NH $\sigma$ and $\sigma^*$ orbitals. 
%SA-CASSCF averaged over 3 states with equal weights.
%The $C_s$ symmetry was imposed for all calculations.
%Note that while S0 and $^1\pi\pi^*$ have an $A'$ symmetry, $^1\pi\sigma^*$ has an $A''$ symmetry.

%Figure~\ref{fig:anl_pes} summarizes SA(3)- and SS-CASSCF PESs along the N--H dissociation. We can see that, at short N--H distances, both SA(3) and SS $^1\pi\sigma^*$ curves exhibits a quasi-bound minima.
%As displayed in Figure~\ref{fig:anl_orbs}, the quasi-bound minimum is associated with a strong Rydberg character of the $\sigma^*$ orbital.
%As the N--H bond length increases, this orbital evolves into a valence character, the $^1\pi\sigma^*$ becomes dissociative accordingly.
%The barriers separating the quasi-bound and dissociative components of SA(3) and SS PESs are 0.58 and 0.49 eV, respectively.
%These values are consistent with the EOM-CCSD prediction \cite{wang2013quantum}. At long N--H distances, the $^1\pi\sigma^*$ curve crosses the S0 curve.
%In SA-CASSCF, the $^1\pi\sigma^*$ curve is entirely below the $^1\pi\pi^*$ curve, and there is no crossing between them. This is also consistent with previous SA-CASSCF reports \cite{sala2014new, ray2018conical}.
%In contrast, as shown in the right inset of Figure~\ref{fig:anl_pes}, SS-CASSCF curves exhibit a crossing at 0.95\r{A}, which is close to the $^1\pi\sigma^*$ local minimum.
%Due to the lack of dynamical correlation, our SS-CASSCF crossing point is quite different from multi-reference perturbation theories \cite{sala2014new, ray2018conical}.
%In general, although our SS-CASSCF is unable to quantitatively predict the crossing point, it can qualitatively explore the crossing that cannot be seen using SA-CASSCF.

\begin{figure}[t]
  \includegraphics[width=8cm,]{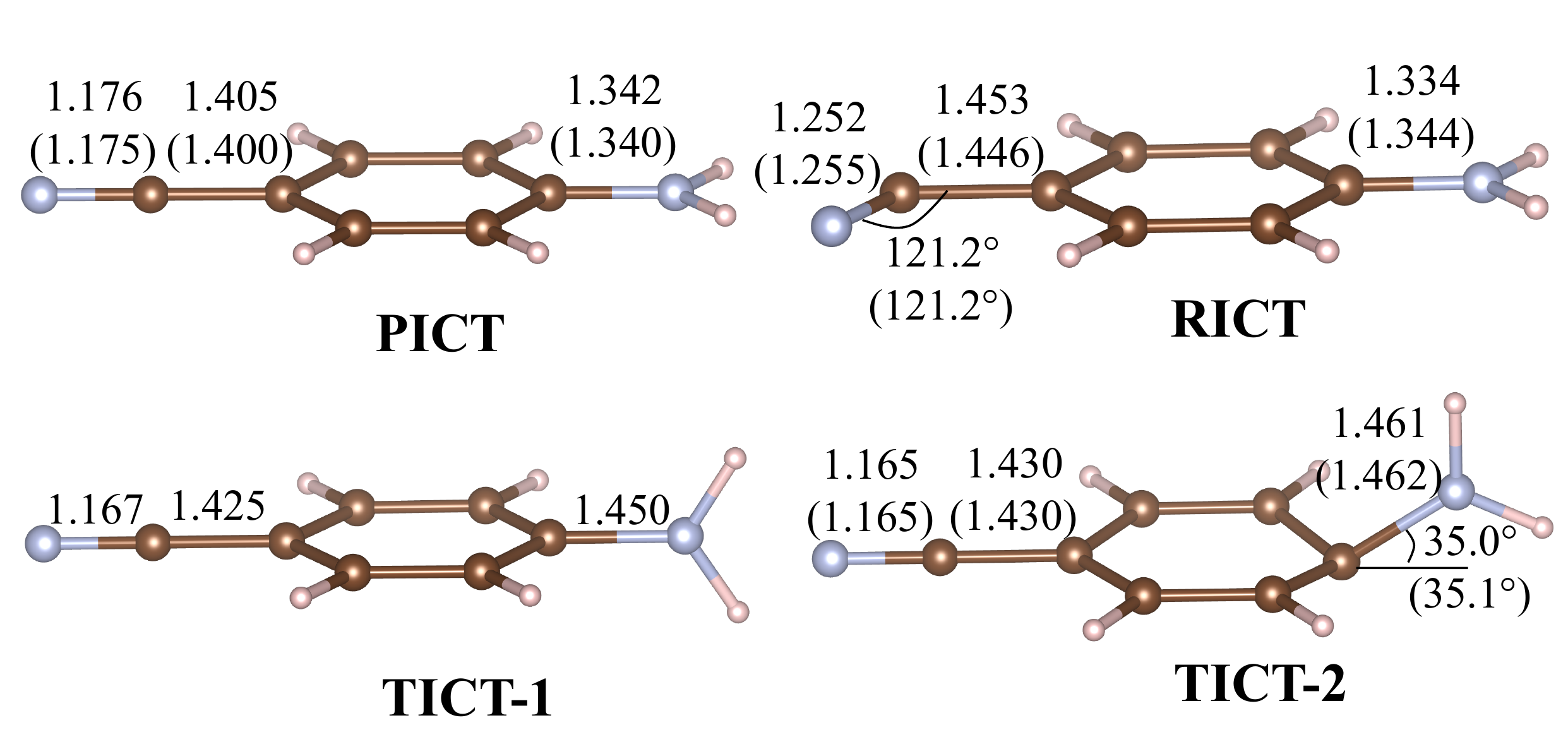}
  \caption{
  SS-CASSCF local minima on the ICT surface of ABN, with corresponding
  SA-CASSCF values from Segado and co-workers \cite{segado2016intramolecular}
  given in parentheses.
  Both approaches used the cc-pVDZ basis set and a
  (12e,11o) active space that contains
  the benzene $\pi$ and $\pi^*$ orbitals,
  the amino nitrogen lone pair,
  and the four $\pi$ and $\pi^*$ orbitals of the cyano group.
  }
  \label{fig:abn_geom}
\end{figure}

Finally, we consider intramolecular charge transfer (ICT) in 
4-aminobenzonitrile (ABN), where Segado and co-workers have shown
\cite{segado2016intramolecular}
that SA-CASSCF predicts a qualitatively incorrect structure
for the minimum-energy geometry of the ICT state.
In addition to a normal fluorescence band associated with a
locally-excited (LE) state, UV excitation of this molecule also
produces an anomalous fluorescence band due to emission from
the ICT state. \cite{grabowski2003,misra2018intramolecular}
%The mechanism of such a dual flouresence in ABN and its family has been extensively studied.
%It is now generally accepted that while the normal band is generated by a locally excited (LE) state, the anomalous one is due to the emission from an ICT state.
%Numerous effors have been devoted to explore the molecular structure of ICT states as well as ICT reaction pathways.
As shown in Figure \ref{fig:abn_geom}, there are multiple local
minima on the ICT surface that could in principle be relevant.
In the planar (PICT) geometry, the amino group, the benzene ring, and the cyano group
all lie in the same plane. \cite{druzhinin2006dynamics}
In the rehybridized (RICT) geometry, the cyano group is wagged in the plane of the ring
due to a rehybridization of the cyano carbon atom from $sp$ to $sp^2$.
\cite{sobolewski1996promotion, sobolewski1996charge}
In the twisted (TICT) geometries, the amino group is rotated so that its plane
is (almost) perpendicular to that of the ring, \cite{grabowski2003,rotkiewicz1973}
and may (TICT-1) or may not (TICT-2) involve the group's bond to the ring 
lying in the ring's plane. \cite{segado2016intramolecular}
While experiment and high-cost post-CASSCF methods agree that the most
stable geometry is twisted, \cite{grabowski2003,misra2018intramolecular}
Segado and co-workers have shown that SA-CASSCF instead predicts the
fully planar geometry to be more stable by almost 15 kcal/mol.
\cite{segado2016intramolecular}
As in aniline, this system involves states for which the
appropriate post-excitation orbital relaxations differ significantly,
and so it is again worth asking whether qualitatively correct
predictions really do require expensive post-CASSCF methods
or simply that each state be able to enjoy molecular orbitals
that have been optimized to suit its needs.

%proposed a new twisted configuration that involves a bending of the amino group relative to the benzene ring \cite{segado2016intramolecular}. However, they found that SA-CASSCF geometry optimization gives this twisted minimum higher than PICT. The correct ordering was only achieved when XMS-CASPT2, which is much more expensive than SA-CASSCF, is employed. 

%In this work, we have optimized several ICT configurations of ABN using our SS-CASSCF. We have used the cc-pVDZ basis set and an active space of 12 electrons in 11 orbitals including the benzene $\pi$ and $\pi^*$ orbitals, the amino nitrogen lone pair, and the four $\pi$ and $\pi^*$ orbitals of the cyano group. Note that our basis set and active space are the as same that in Ref.~\onlinecite{segado2016intramolecular}. We have also examined two twisted structures: without and with the bending of amino group. We denote these structure as TICT-1 and TICT-2, respectively.

\begin{table}[t]
  \normalsize
  \caption{
           \label{tab:abn_energy}
           Energy differences $\Delta E$ relative to the S0 minimum (in kcal/mol)
           and dipoles (in Debye) for ABN's ICT state at different minima on its
           PES for both SS-CASSCF and the 3-state SA approach of
           Segado and coworkers. \cite{segado2016intramolecular}\\
           % \normalsize SS- and SA(3)-CASSCF relative energies $\Delta E$ (in kcal/mol) to the S0 minimum and dipole moments $\mu$ (in Debye) of different excited states of ABN. The SA(3)-CASSCF data were taken from Ref.~\onlinecite{segado2016intramolecular}\\
          }
  \begin{ruledtabular}
  \begin{tabular}{ccccccccccc}
     \multirow{2}{*}{States} && \multicolumn{2}{c}{SS-CASSCF} & &\multicolumn{2}{c}{SA(3)-CASSCF \cite{segado2016intramolecular}} \\
    \cline{3-4} \cline{6-7}
           && $\Delta E$  & $\mu$  &    &$\Delta E$  &$\mu$  \\
    \hline
    S0     && 0           &6.08      &  &0          &5.40 \\
    PICT   && 139.19      &11.92     &  &146.15     &11.23 \\
    RICT   && 140.50      &13.95     &  &152.77     &10.45 \\
    TICT-1 && 131.70      &13.66     &  & -         &- \\
    TICT-2 && 128.02      &10.66     &  &160.93     &10.76 \\
  \end{tabular}
  \end{ruledtabular}
\end{table}

\begin{figure}[t]
  \includegraphics[width=8cm,]{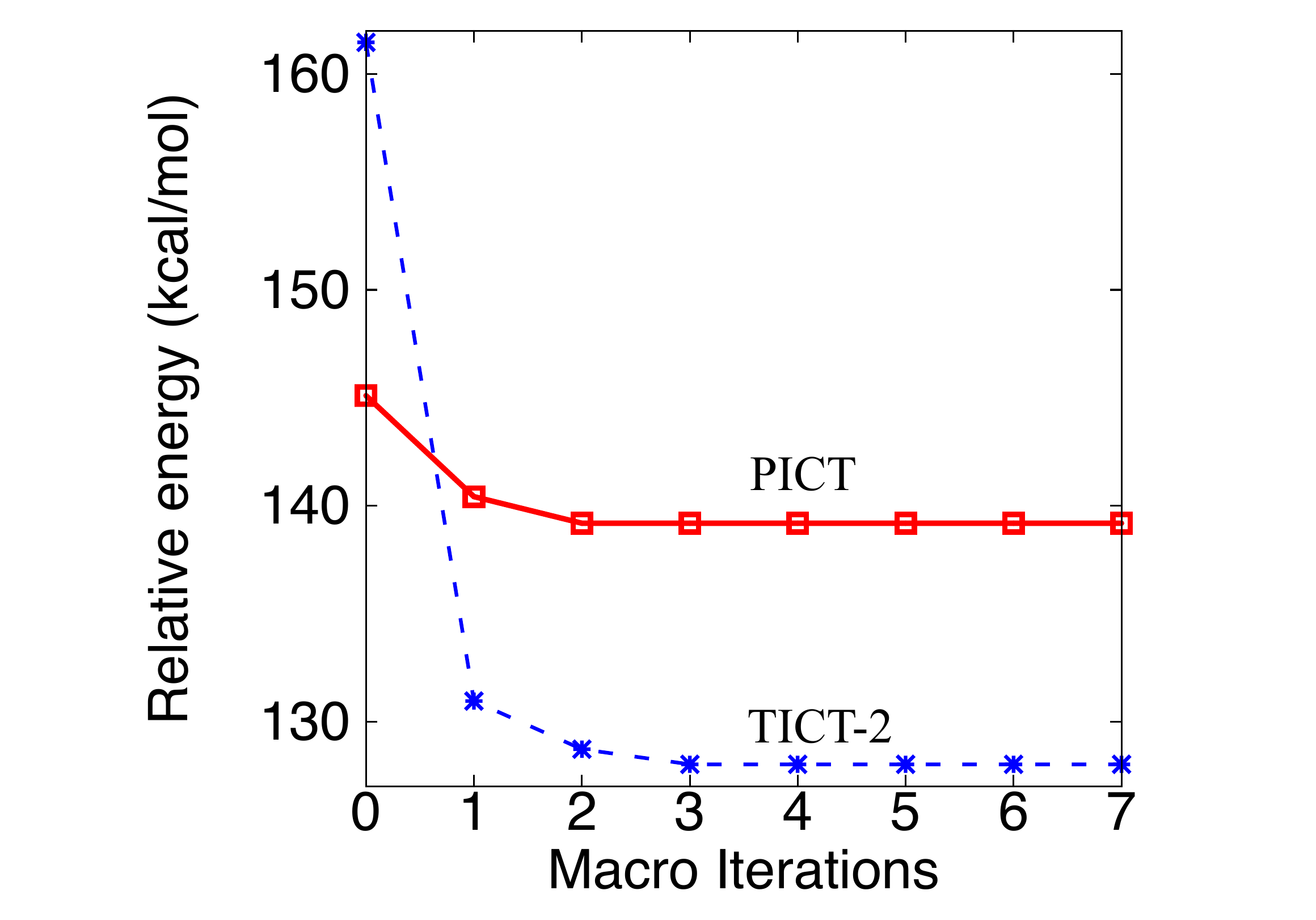}
  \caption{As an example of how much the SA bias against ABN's
           ICT state varies with geometry,
           we plot the lowering of this state's energy during SS-CASSCF
           optimizations that start from the SA-CASSCF wave functions
           at both the SS-CASSCF planar (PICT) and twisted (TICT-2)
           geometry.
           Energies are reported relative to the $S0$ minimum.
          }
  \label{fig:abn_iter}
\end{figure}

\begin{figure}[t]
  \includegraphics[width=8cm,]{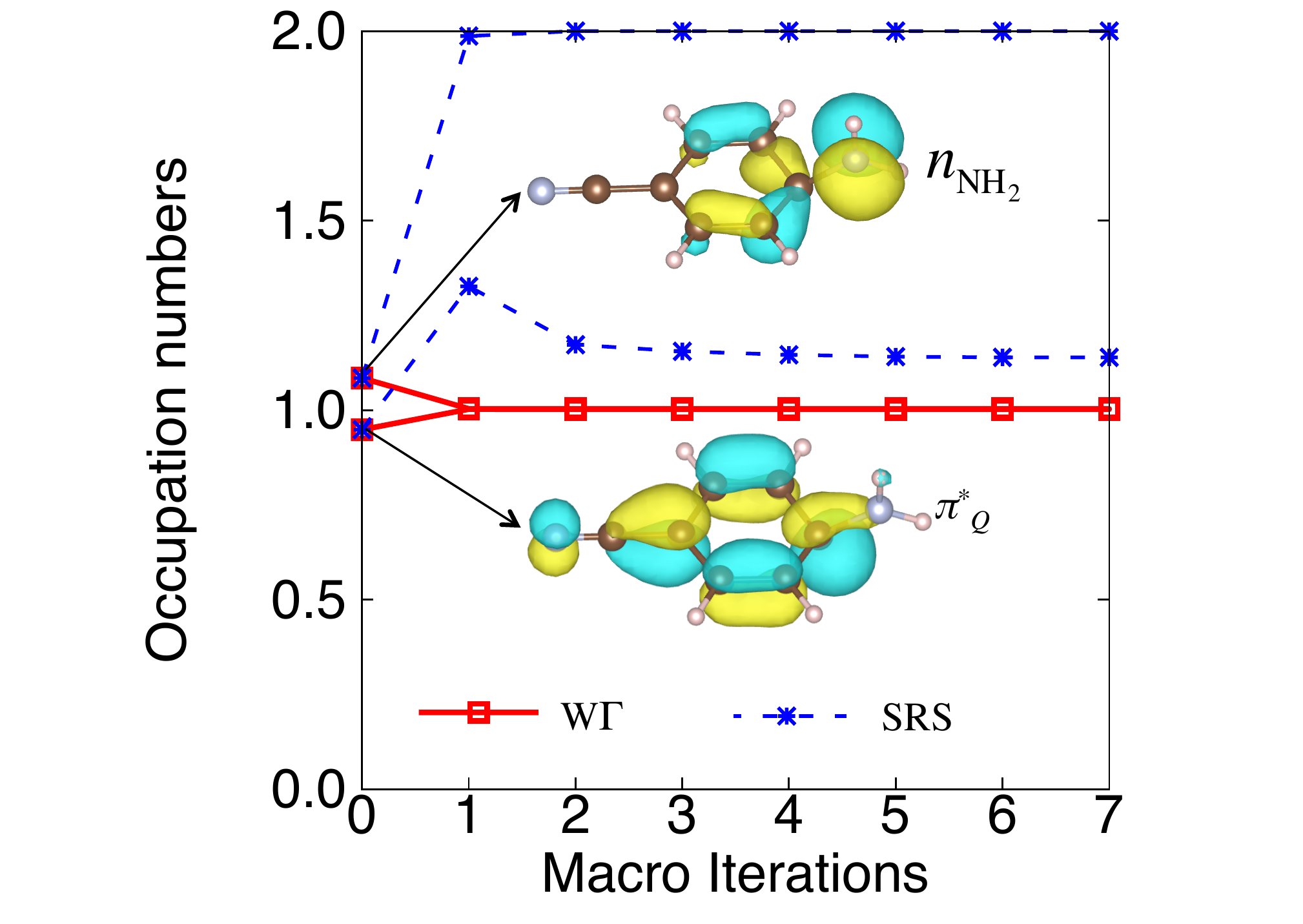}
  \caption{Orbital occupation changes during SS-CASSCF's
           wave function optimization of ABN's ICT state at the TICT-2 geometry,
           with SA-CASSCF used as the guess.
           Our $W\Gamma$ approach maintains the correct state character,
           \cite{segado2016intramolecular}
           whereas simple root selection (SRS), which selects the CASCI root
           based on energy ordering, collapses to a non-charge-transfer state.
           %The insets are shapes of $n_{\text{NH}_2}$ and $\pi^*_{Q}$ orbitals.
          }
  \label{fig:abn_occ}
\end{figure}

Although Figure \ref{fig:abn_geom} reveals that the structures predicted
by SS-CASSCF agree with those from SA-CASSCF, Table \ref{tab:abn_energy} shows
that the energetics are quite different and that, as anticipated,
state-specific orbital relaxation leads to predictions that agree
qualitatively with experiment and post-CASSCF methods.
In particular, SS-CASSCF agrees with XMS-CASPT2 in predicting that TICT-2
is the most stable ICT structure. \cite{segado2016intramolecular}
By allowing orbitals to fully relax following the charge transfer,
SS-CASSCF stabilizes the ICT state at all geometries,
but the effect is much stronger in the twisted geometries, as shown
for example in Figure \ref{fig:abn_iter}.
These results imply that the SA compromise for orbital shapes,
which is always worrisome when the non-charge-transfer states
(S0, LE) outnumber charge transfer states (ICT),
creates a significantly stronger bias against the ICT state at some
geometries than at others, ruining the balance that SA is in
principle supposed to provide.
In contrast, our root-tracking method's ability to tailor orbitals
for individual states without losing track of those states during the wave
function optimization (see Figure \ref{fig:abn_occ}) allows for
the energetically significant re-polarizations that all orbitals
are expected to undergo following a charge transfer.
Happily, capturing these effects is sufficient to bring the prediction
in line with experiment, again suggesting that one can go a long way
in repairing the failures of SA-CASSCF without resorting to the
expense of post-CASSCF methods.
When one considers that this should be especially true in charge
transfer systems and that, in order to separate charge over a significant
distance for technological purposes, these systems often contain dozens
or even hundreds of atoms, the advantages of improving accuracy without
increasing cost become even more desirable.

{\it Conclusion.}
In summary, we have shown that by providing each excited state with orbitals
that are optimal for its needs, qualitative failures in potential energy surfaces
can be corrected without significantly increasing computational cost.
Moreover, state-specific orbitals dramatically simplify the evaluation
of analytic gradients by eliminating the wave function response calculations
that are required by state averaged approaches.
In particular, we have shown that this state-specific approach, in which
collapse to other states is avoided via novel root tracking,
makes qualitative improvements over state averaging in a C--S double
bond dissociation, in an amino N--H dissociation,
and in an intramolecular charge transfer geometry relaxation.
Our approach's general nature --- it requires the same basic ingredients
as can be found in ground state CASSCF implementations --- should allow these
advantages to be enjoyed across a wide variety of excited state
applications, with particular promise for cases like charge transfer
and core excitation that involve substantial post-excitation orbital relaxation.

Looking forward, there are a number of clear ways in which the methodology
can be enhanced to further widen its utility.
Thanks to its two-step optimization formulation, it should be quite straightforward
to tackle large active spaces with selective CI or the density matrix
renormalization group.
Combining these approaches with excited-state-specific orbital relaxations looks
especially promising for metal-ligand charge transfer complexes, where
double-d-shell effects and large ligand $\pi$ systems quite rapidly push
the desired active space beyond the reach of conventional solvers.
A second obvious priority is enabling inter-state properties like transition
dipoles and derivative couplings.
Although our approach leads naturally to a situation in which different states
are expressed in different molecular orbital bases, making inter-state matrix
elements less straightforward, non-orthogonal configuration
interaction techniques can be adapted to meet this challenge.
These same techniques are also relevant for re-diagonalizing a pair of
states near a conical intersection, where the lack of strict orthogonality
within our state-specific approach becomes a real concern.
With these various improvements, it should be possible to bring the benefits
of state-specific orbital optimization, and the qualitative improvements in accuracy
that it offers, to the wide array of property predictions on which
spectroscopists depend.

{\it Acknowledgement.}
This work was supported by the Early Career Research Program
of the Office of Science, Office of Basic Energy Sciences,
the U.S. Department of Energy, grant No.\ {DE-SC0017869}.
Calculations ran on the Berkeley Research
Computing Savio cluster.

\bibliography{main}
\end{document}